\documentclass[14pt]{article}

\usepackage{amsfonts}
\usepackage{epsfig}
\usepackage{a4}
\usepackage{theorem}
\usepackage{amsmath}
\usepackage{amssymb}
\usepackage{graphicx}
\usepackage{slashed}
\usepackage{pstricks}
\usepackage{color}

\newcommand{\be}[1]{\begin{equation}\label{#1}}
\newcommand{\ee}{\end{equation}}
\newcommand{\ba}[1]{\begin{eqnarray}\label{#1}}
\newcommand{\ea}{\end{eqnarray}}
\makeatother
\date{}
\begin{document}
\title{COSMOLOGICAL CORRELATORS, IN-IN FORMALISM AND DOUBLE COPY}
\author{
A.R.~Fazio 
\vspace{3mm} \\
\small Departamento de F\'isica\\
\small Ciudad Universitaria\\
\small Bogot\'a, D.C. Colombia\\
\small arfazio@unal.edu.co\vspace{3mm}\\}
\maketitle
\begin{abstract}
We are investigating if the double copy structure as product of scattering amplitudes of gauge theories applies to cosmological correlators computed in a class of theories for inflation by the operatorial version of the In-In formalism of Schwinger-Keldysh. We consider tree level momentum-space correlators involving primordial gravitational waves with different polarizations and the scalar curvature fluctuations on a three dimensional fixed spatial slice. The correlators are sum of terms factorized in a time dependent scalar factor, which takes into account the curved background where energy is not conserved, and in a so-called tensor factor, constructed by polarization tensors. In the latter we recognize scattering amplitudes in four dimensional Minkowski space spanned by three points gravitational amplitudes related by double copy to those of gauge theories. Our study indicates that gravitational waves are double copy of gluons and the primordial scalar curvature is double copy of a scalar with Higgs-like interactions.
\end{abstract}

\section{Theories of inflation}	

The theories of inflation considered in this research are in the class of the so called single field slow roll inflation. They contain a real scalar field $\phi(\vec{x},t)$ in the $\zeta$- gauge of Maldacena \cite{Malda},\cite{Wein} with no quantum fluctuations around its non zero, homogeneous and time dependent vacuum expectation value $\bar{\phi}(t)$, and gravitational interactions in a background space-time whose spatial part of the metric has the form
\begin{equation}
g_{ij}=a^2(t)\exp(2\zeta)\left[\exp\gamma\right]_{ij},\,\,\,\,\,\,\gamma_{ii}=0,\,\,\,\,\, \partial_i\gamma_{ij}=0.
\label{metric}
\end{equation}  
$a(t)$ is the Robertson-Walker scale factor describing the expansion of the background and $\gamma_{ij}(\vec{x},t)$ is a gravitational amplitude. Three-vectors are intended to be such, relative to the spatial metric (\ref{metric}). The so-called primordial curvature perturbation, $\zeta(\vec{x},t)$, is a scalar field which has a constant value outside the horizon meaning at the epoch  with temperatures of $T\sim 10^{-1} MeV$ \cite{TASI2012}. This constant value determines the total energy density perturbation and it provides the main initial condition for the subsequent evolution of all perturbations. Some proof of the conservation at tree level can be found in \cite{Malda},  arguments for constancy at one-loop level are provided in \cite{Pimentel-Senatore 12036651}. It is actually a current research topic to prove the constancy of $\zeta$ to all order of perturbation theory. One hope related to our research article is that it could help in this effort by the methods we are going to present.

The others metric tensor components are given in the Arnowitt-Deser-Misner (ADM) formalism   
\begin{equation}
g_{00}= -N^2+g_{ij}N^i N^j,\,\,\,\,\  g_{i0}=g_{ij}N^j.
\label{metric1}
\end{equation} 
From the three dimensional metric $g_{ij}$ one calculates the spatial Ricci tensor $R^{(3)}_{ij}$ and the Ricci scalar intrinsic curvature $R^{(3)}=g^{ij}R^{(3)}_{ij}$. The corresponding extrinsic curvature tensor is
\begin{equation}
K_{ij}=-\frac{1}{2N}\left(\dot{g}_{ij}-\nabla_iN_j-\nabla_j N_i\right).
\end{equation}
 To study correlators between graviton fluctuations and primordial curvature we can solve the constraints equations for the lapse and shift functions and truncate the solutions to
\begin{equation}
N=1+\frac{\dot{\zeta}}{H}\,\,\,\,\,\,\,\,\,\,\,\,\,\,\,\,\,\, N_i=-\frac{1}{Ha^2}\partial_i\zeta +\epsilon\partial_i\frac{1}{\nabla^2}\dot{\zeta},
\end{equation} 
written in terms of the Hubble rate $H$ and of a slow roll parameter $\epsilon$ 
\begin{equation}
H=\frac{\dot{a}}{a}\,\,\,\,\,\,\,\,\,\,\, \epsilon= -\frac{\dot{H}}{H^2}=\frac{\dot{\bar{\phi}}^2}{2H^2}\,\,\,.
\end{equation}
To perturbatively compute the quantum correlators the needed fields in the interaction picture are
\begin{equation}
\zeta(\vec{x},t)= \int d^3 q\left[\alpha(\vec{q})\zeta_q(t)\exp(i\vec{q}\cdot \vec{x})+\alpha^\dag(\vec{q})\zeta_q^*(t)\exp(-i\vec{q}\cdot \vec{x})\right]
\end{equation}
\begin{eqnarray}
&&\gamma_{ij}(\vec{x},t)=\int d^3 q\sum_{\lambda=\pm 2}\nonumber\\&&\left[\exp(i\vec{q}\cdot \vec{x})e_{ij}(\hat{q},\lambda)\alpha(\hat{q},\lambda)\gamma_q(t)+ \exp(-i\vec{q}\cdot \vec{x})e^*_{ij}(\hat{q},\lambda)\alpha^\dag(\hat{q},\lambda)\gamma^*_q(t)\right],
\end{eqnarray}
being $\lambda$ the helicity and $e_{ij}(\hat{q},\lambda)$ is a polarization tensor while $\alpha(\vec{q})$ and $\alpha(\hat{q},\lambda)$ are annihilation operators with algebra
\begin{equation}
\left [\alpha(\vec{q}), \alpha^\dag(\vec{q}\,\,')\right]=\delta^3(\vec{q}-\vec{q}\,\,')\,\,\,\,\,\,\,\, \left [\alpha(\vec{q}), \alpha(\vec{q}\,\,')\right]=0
\end{equation}
\begin{equation}
\left [\alpha(\vec{q}\,'\lambda), \alpha^\dag(\vec{q}\,\,' , \lambda')\right]=\delta_{\lambda\lambda'}\delta^3(\vec{q}-\vec{q}\,\,')\,\,\,\,\,\,\,\, \left [\alpha(\vec{q},\lambda), \alpha(\vec{q}\,',\lambda')\right]=0.
\end{equation}
Moreover $\zeta_q(t)$ and $\gamma_q(t)$, with $q=|\vec{q}|$, are suitable normalized solutions of the equations
\begin{equation}
\frac{d^2}{dt^2}\zeta_q(t)+\left[\frac{d\log\left(a^3\epsilon\right)}{dt}\right]\frac{d}{dt}\zeta_q(t) +\frac{q^2}{a^2 }\zeta_q(t)=0;
\label{MUK}
\end{equation}

\begin{equation}
\frac{d^2}{dt^2}\gamma_q(t)+3H\frac{d}{dt}\gamma_q(t) +\frac{q^2}{a^2 }\gamma_q(t)=0,
\label{GRAV}
\end{equation}

obtained by using the metric prescriptions (\ref{metric}) and (\ref{metric1}) in the Einstein Hilbert action in the ADM parametrization and taking the quadratic part

\begin{eqnarray}
&& S_2=-\frac{1}{8}\int d^3x dt\,\, a(t)\partial_m\gamma_{ij}\partial_m\gamma_{ij} + \frac{1}{8}\int d^3x dt\,\, a^3(t)\partial_t\gamma_{ij}\partial_t\gamma_{ij}\nonumber\\
&& - \int d^3x dt\,\left\{a\epsilon(\nabla\zeta)^2 +\frac{d}{dt}\,\,(a^3\epsilon)\zeta\frac{\partial}{\partial t}\zeta + a^3\epsilon \zeta \frac{\partial^2}{\partial t^2}\zeta\right\}.
\end{eqnarray}
The space integral is on $\mathbb{R}^3$ and the indices contractions are made by the unit metric $\delta_{ij}$. The two initial conditions needed to solve (\ref{MUK}) and (\ref{GRAV}) are related to the fact that at early times the Hubble expansion is negligible and at late times $\zeta_q(t)$ as well as $\gamma_q(t)$ go to the constant value taken outside the horizon.

\section{Correlator $<\zeta\gamma\gamma>$}

We start by calculating the tree-level contribution to the correlation function of one $\zeta$ field and two gravitational waves $\gamma$ fields. This correlator is available in literature \cite{Pimentel-thesis},\cite{malda-pimentel},\cite{Gates},\cite{Farrow},\cite{quadri}, however we will write it here in a different form with the aim of distinguishing between a scalar and tensor part at integral level and on the latter look for a connection with a scattering amplitudes in four dimensional Minkowski space gauge theories (double copy).
\begin{eqnarray}
&&G(\vec{p}_1, \vec{p}_2,\vec{p}_3)_{ij;lm}\equiv \int d^3x_1 d^3x_2d^3x_3 \exp(i\vec{p}_1\cdot \vec{x}_1)\exp(i\vec{p}_2\cdot \vec{x}_2)\exp(i\vec{p}_3\cdot \vec{x}_3)\nonumber\\
&&\times \langle\zeta(\vec{x}_1,t)\gamma_{ij}(\vec{x}_2,t)\gamma_{lm}(\vec{x}_3,t)\rangle
\label{ilprimo}
\end{eqnarray}
In the so-called In-In formalism \cite{Wein},\cite{Atlad}, suitable for computing correlations of fluctuations at a given time, the lowest perturbative order is equal to
\begin{eqnarray}
&&G(\vec{p}_1, \vec{p}_2,\vec{p}_3)_{ij;lm}=i\int_{-\infty}^t dt_1 \int d^3x_1 d^3x_2d^3x_3 <[H_{\zeta\gamma\gamma}(t_1), \nonumber\\
&&\exp(i\vec{p}_1\cdot \vec{x}_1)\exp(i\vec{p}_2\cdot \vec{x}_2)\exp(i\vec{p}_3\cdot \vec{x}_3)\zeta(\vec{x}_1,t)\gamma_{ij}(\vec{x}_2,t)\gamma_{lm}(\vec{x}_3,t) ]>.\nonumber\\
\label{ININ1}
\end{eqnarray}
The time dependent interaction Hamiltonian, which includes only interactions for which the time integral in (\ref{ININ1}) is convergent as $t\rightarrow\infty$, amounts to \cite{Wein}
\begin{equation}
H_{\zeta\gamma\gamma}(t)\equiv A(t) +\frac{d}{d t}B(t)
\label{zgg}
\end{equation}
where

\begin{equation}
A(t)=- \frac{\epsilon(t) a(t)}{8}\int \zeta (\vec{x},t)\partial_m\gamma_{ri}(\vec{x},t)\partial_m\gamma_{ri}(\vec{x},t) d^3 x,
\label{A-prima}
\end{equation}

\begin{equation}
B(t)=\frac{a(t)}{8H(t)}\int \zeta(\vec{x},t)\partial_m\gamma_{ri}(\vec{x},t)\partial_m\gamma_{ri}(\vec{x},t) d^3 x.
\end{equation}

The calculation of the commutators in (\ref{ININ1}) is performed by using 
\begin{equation}
\left[\zeta(\vec{x},t), \zeta(\vec{x}\,', t')\right] =\int d^3p\, e^{i\vec{p}\cdot (\vec{x}-\vec{x}\,')} (\zeta_p(t)\zeta^*_p(t')-\zeta_p^*(t)\zeta_p(t'))
\label{Green1}
\end{equation} 
as well as

\begin{equation}
\left[\gamma_{ij}(\vec{x},t), \gamma_{\ell m}(\vec{x}\,', t')\right] =\int d^3p\,\, e^{i\vec{p}\cdot (\vec{x}-\vec{x}\,')} \Pi_{ij;lm}(\hat{p})(\gamma_p(t)\gamma^*_p(t')-\gamma_p^*(t)\gamma_p(t'))
\label{Green2}
\end{equation} 
where
\begin{equation}
\Pi_{ij;lm}(\hat{p})=\sum_\lambda e_{ij}(\hat{p},\lambda)e^{*}_{lm}(\hat{p},\lambda).
\end{equation}
The chosen polarization tensors are symmetric, transverse and traceless
\begin{equation}
e_{ij}(\hat{p},\lambda)=e_{ji}(\hat{p},\lambda)\,\,\,\,\,\,\, p_i e_{ij}(\hat{p},\lambda)=0\,\,\,\,\,\,\,\ e_{ii}(\hat{p},\lambda)=0.
\label{gaugepol}
\end{equation}
By the described operatorial method \cite{Wein},\cite{Ashead} we get
\begin{eqnarray}
&&G(\vec{p}_1, \vec{p}_2,\vec{p}_3)_{ij;lm}\equiv\delta^3 (\vec{p}_1+\vec{p}_2+\vec{p}_3)G_1(p_1,p_2,p_3)\times\nonumber\\
&&(2\vec{p}_2\cdot\vec{p}_3)\Pi_{ij;rn}(\hat{p}_2)\Pi_{rn;lm}(\hat{p}_3)
\label{G}
\end{eqnarray}
and the scalar factor is
\begin{eqnarray}
&& G_1 (p_1,p_2,p_3)=\frac{i (2\pi)^{12}}{8}\int_{-\infty}^t d t_1 \epsilon(t_1) a(t_1)\left[\zeta_{p_1}(t_1)\zeta^*_{p_1}(t)\gamma_{p_2}(t_1)\gamma^{*}_{p_2}(t)\gamma_{p_3}(t_1)\gamma^{*}_{p_3}(t)+\right.\nonumber\\
&&\left.-\zeta_{p_1}(t)\zeta^*_{p_1}(t_1)\gamma_{p_2}(t)\gamma^{*}_{p_2}(t_1)\gamma_{p_3}(t)\gamma^{*}_{p_3}(t_1)\right].
\label{G1}
\end{eqnarray}
The appearing of the delta for the three-momentum conservation is due to the time dependent background.
The time independent tensor factor in (\ref{ilprimo}) is
\begin{equation}
2\vec{p}_2\cdot\vec{p}_3\Pi_{ij;rn}(\hat{p}_2)\Pi_{rn;lm}(\hat{p}_3),
 \end{equation}
which after being saturated with specific gravitons polarization tensors of given helicities amounts
\begin{equation}
4(2\vec{p}_2\cdot\vec{p}_3)\epsilon_{ij}(\hat{p}_2,\lambda)\epsilon_{ij}(\hat{p}_3,\lambda').
\label{project}
\end{equation}
The transversality and traceless of the polarization tensors (\ref{gaugepol}) has been used to deduce [\cite{Weinbook}]
\begin{eqnarray}
&&\Pi_{ij;kl}(\hat{q})= \delta_{ik}\delta_{jl}+\delta_{il}\delta_{jk}-\delta_{ij}\delta_{kl} +\delta_{ij}\hat{q}_k\hat{q}_l +\delta_{kl}\hat{q}_i\hat{q}_j -\delta_{ik}\hat{q}_j\hat{q}_l\nonumber\\
&& -\delta_{il}\hat{q}_j\hat{q}_k -\delta_{jk}\hat{q}_i\hat{q}_l-\delta_{jl}\hat{q}_i\hat{q}_k +\hat{q}_i\hat{q}_j\hat{q}_k\hat{q}_l,
\label{sumpol}
\end{eqnarray}
which allowed the projections in (\ref{project}). We are going to use a new set of helicity spinors variables allowing to easily make connection with gauge theory amplitudes in four dimensions.
\section{$SO(3)$ spinor helicity formalism for inflation}
In order to explore the properties of factorization by double copy we introduce a spinor helicity formalism for inflation. We shall follow a different approach than in \cite{Malda},\cite{kltINFLA}, meaning that our spinors are not based on those of the representations $D^{(0,\frac{1}{2})}$, $D^{(\frac{1}{2},0)}$ of the ortho-chronus proper Lorentz group. Our spinors will be those to construct the irreducible representations of $SO(3)$, the group of invariance of the dynamics of fluctuations in the considered class of theory \cite{Moschella}. For the construction of the above spinor variables we follow \cite{Fogli},\cite{Cartan}.

A controvariant spinor of lowest dimension of $SO(3)$ is 
\begin{equation}
\xi=\begin{pmatrix}
  \xi^{1} \\
  \xi^{2} \\
\end{pmatrix}
\label{xi}
\end{equation}

which transforms as 
\begin{equation}
\xi ' =u\xi
\label{T1}
\end{equation}
being $u\in SU(2)$.

We may construct invariants by introducing the covariant spinor $\eta$ which transforms as
\begin{equation}
\eta' =\eta u^{-1}=\eta u^\dag,
\end{equation}
and the consequent $SU(2)$ invariant scalar product
\begin{equation}
\eta\xi = \eta'\xi'=\eta_i\xi^i.
\end{equation}

The phase convention is fixed as
\begin{equation}
\xi_i\equiv\xi^{i *}.
\label{phase}
\end{equation}

Consider the four component tensor $\Xi^j _i\equiv\xi^j\xi_i$ which can be reduced to the following irreducible three components object
\begin{equation}
\hat{\Xi}^j_i\equiv  \xi^j\xi_i -\frac{1}{2}\delta^j_i \xi^k\xi_k
\label{3-vec}
\end{equation}
equivalent to a real three vector $\vec{p}$ by
\begin{equation}
\hat{\Xi}= \vec{\sigma}\cdot \vec{p}=\begin{pmatrix} 
p_3 & p_1-\imath p_2 \\
p_1+\imath p_2  & -p_3\\
\end{pmatrix},
\end{equation}
due to the $SU(2)$ transformation property
\begin{equation}
\hat{\Xi}\rightarrow \hat{\Xi'}=u\hat{\Xi}u^\dag.
\end{equation}

By inverting the equation (\ref{3-vec}) the three vector $\vec{p}$ can be written as bi-spinor as follows
\begin{equation}
\vec{p}=\frac{1}{2}\xi^\dag\vec{\sigma}\xi
\end{equation}
which we rewrite as
\begin{equation}
\vec{p}=\frac{1}{2}\xi_p^\dag\vec{\sigma}\xi_p
\end{equation}
and say that the spinor $\xi_p$ is the ``root square" of the vector $\vec{p}$. Finally it is useful the following notation
\begin{equation}
p^j_i\equiv  \xi_p^j\xi_{pi} -\frac{1}{2}\delta^j_i (\xi_p, \xi_p),
\end{equation}
where
\begin{equation}
\left(\xi_p,\xi_q\right)=\xi_p^\ell\xi_{q\,\ell}.
\end{equation}
Since the square of the matrix $p^j_i$ is the unit matrix times the square of the vector $\vec{p}$ it is deduced that, due to the phase convention (\ref{phase}),
\begin{equation}
(\xi_p, \xi_{p}) = 2 |\vec{p}|
\end{equation}
and consequently
\begin{equation}
p^j_i\equiv  \xi_p^j\xi_{pi} -|\vec{p}|\,\delta^j_i.
\end{equation}

For analogous reasons the scalar product 
\begin{equation}
\vec{p}\cdot \vec{q}= \frac{1}{2}|\xi_{p\,k}\xi_q^k|^2- \frac{1}{4} \xi_{p\,k}\xi_p^k \xi_{q\,k}\xi_q^k =  \frac{1}{2}|(\xi_{p},\xi_q)|^2-|\vec{p}||\vec{q}|.
\end{equation}
A momentum conservation condition $\vec{p}_1 +\vec{p}_2+\vec{p}_3 =0$ is written in terms of spinors
\begin{equation}
\xi_{p_1}^j\xi_{{p_1}i} + \xi_{p_2}^j\xi_{{p_2}i}+ \xi_{p_3}^j\xi_{{p_3}i}= \delta^j_i (|\vec{p}_1|+|\vec{p}_2|+ |\vec{p}_3| ).
\label{cons}
\end{equation}
We can contract (\ref{cons})  with $\xi_{p_1\,j}$ and $\xi_{p_2}^i$ to find
\begin{equation}
\xi_{p_1\, k}\xi_{p_3}^k\xi_{p_3\ell}\xi_{p_2}^\ell=(|\vec{p}_3|-|\vec{p}_1|-|\vec{p}_2|)\xi_{p_1\,k}\xi_{p_2}^k
\end{equation}
and analogously 
\begin{equation}
\xi_{p_2}^\ell (-\imath\sigma_2)_{\ell j}\xi_{p_1}^j\xi_{p_1\, i} (-i\sigma_2)^{in}\xi_{p_3\, n} = {\xi_{p_3}}_k \xi_{p_2}^k (|\vec{p}_1|+|\vec{p}_2| + |\vec{p}_3| )
\label{cons*}
\end{equation}
\begin{equation}
\xi_{p_2}^\ell{(-\imath\sigma_2)}_{\ell j}\xi_{p_1}^j\xi_{p_2\,m}(-\imath\sigma_2)^{m\,n}\xi_{p_1\,n} =-|\vec{p}_3|^2 + \left(|\vec{p}_2| - |\vec{p}_1|\right)^2.
\label{cons**}
\end{equation}
\subsection{Polarization vectors} 


The three dimensional polarization vectors for helicity $\pm 1$ massles fields are written in the Coulomb gauge in terms of spinors as
\begin{equation}
{\epsilon}(\vec{p},+)^j_i= \frac{\xi_p^j  (-\imath\sigma_2)_{i\ell}\xi_p^{ \ell}}{\sqrt{2}|\vec{p}|}
\label{+}
\end{equation}
\begin{equation}
\epsilon(\vec{p}, -)^j_i=-\frac{(-\imath\sigma_2)^{j\ell}\xi_{p \ell}\xi_{pi}}{\sqrt{2}|\vec{p}|}.
\label{-}
\end{equation}
The (matrices) polarization vectors (\ref{+}) and (\ref{-}) satisfy the following properties:
\begin{equation}
\left({\epsilon}(\vec{p},+)\right)^\dag={\epsilon}(\vec{p},-)
\end{equation}
implying for the corresponding vectors 
\begin{equation}
\vec{\epsilon}(\vec{p},\pm)^{*}=\vec{\epsilon}(\vec{p},\mp).
\end{equation}
For the transversality condition
\begin{equation}
\vec{p}\cdot \vec{\epsilon}(\vec{p},\pm)=0.
\label{per2}
\end{equation}
Observe in fact that for the matrices $A^j_i$ and $B^j_i$ associated respectively to the vectors $\vec{a}$ and $\vec{b}$ the following property holds
\begin{equation}
\{A,B\}=2\vec{a}\cdot\vec{b}\,\, {\rm I}_{2\times 2}
\label{anti}
\end{equation}
and consequently
\begin{equation}
{\rm Tr} AB = 2\vec{a}\cdot\vec{b}. 
\label{trace}
\end{equation}
A straightforward application of (\ref{trace}) allows to prove (\ref{per2}) and
\begin{equation}
\vec{\epsilon}(\vec{p},+)\cdot \vec{\epsilon}(\vec{p},-)^*=\vec{\epsilon}(\vec{p},+)\cdot \vec{\epsilon}(\vec{p},+)=0
\end{equation}  
arising from the general property (\ref{anti})
\begin{equation}
\epsilon(\vec{p},\pm)^j_\ell \epsilon (\vec{p},\pm)^\ell_i = \vec{\epsilon}(\vec{p},\pm)^2\delta^j_i.
\end{equation}
Finally
\begin{equation}
|\vec{\epsilon}(\vec{p},\pm)|^2=1.
\end{equation}

By taking the momentum $\vec{p}=(0,0,p)$, $p>0$, and normalizing the spinor $\xi_p=\begin{pmatrix}
  \sqrt{2p} \\
  0 \\
\end{pmatrix}$
we obtain for the matrix
\begin{equation}
\epsilon(\vec{p},+)^j_i=\begin{pmatrix} 
0 &\sqrt{2} \\
0  & 0\\
\end{pmatrix}=\sqrt{2}{(\sigma_+)}^j_i
 \end{equation}
 which corresponds to the polarization vector 
 \begin{equation}
 \vec{\epsilon}(\vec{p},+)=\frac{1}{\sqrt{2}}\begin{pmatrix}
  1 \\
  \imath \\0
\end{pmatrix}.
\end{equation}.
\subsection{Some expressions involving polarization vectors}
Now let us calculate all possible contractions of polarization vectors and momenta
of different helicities:
\begin{equation}
\vec{\epsilon}(\vec{p},+)\cdot \vec{\epsilon}(\vec{q},+) =-\frac{( \xi_p^i(i\sigma_2)_{i\ell}\xi_q^\ell)^2}{4|\vec{p}||\vec{q}|}\,\,,
\end{equation}

\begin{equation}
\vec{\epsilon}(\vec{p},-)\cdot \vec{\epsilon}(\vec{q},-) =-\frac{( \xi_{p\,i}(i\sigma_2)^{i\ell}\xi_{q\,\ell})^2}{4|\vec{p}||\vec{q}|}\,\, ,
\end{equation}

\begin{equation}
\vec{\epsilon}(\vec{p},+)\cdot \vec{\epsilon}(\vec{q},-) =\frac{( \xi_q,\xi_p)^2}{4|\vec{p}||\vec{q}|}\,\, ,
\end{equation}

\begin{equation}
\vec{q}\cdot \vec{\epsilon}(\vec{p},+)=\frac{1}{2\sqrt{2}|\vec{p}|}(\xi_q,\xi_p) \left(\xi_q^i(-i\sigma_2)_{i\ell}\xi_p^\ell\right)\,\, ,
\end{equation}

\begin{equation}
\vec{q}\cdot \vec{\epsilon}(\vec{p},-)=\frac{1}{2\sqrt{2}|\vec{p}|}(\xi_p,\xi_q) \left(\xi_{p\, i}(-i\sigma_2)^{i\ell}\xi_{q\,\ell}\right).
\end{equation}

\subsection{Massless helicity two wave function}
Due to the gauge conditions (\ref{gaugepol}) and the behavior under rotations around the direction of the momentum  we can take the polarization tensors as direct products of helicity $1$ polarization vectors
\begin{equation}
e_{lm}(\hat{p}, +2) = \epsilon_l(\hat{p}, +) \epsilon_m(\hat{p}, +) 
\label{2+}
\end{equation}
and 
 \begin{equation}
e_{lm}(\hat{p}, -2) = \epsilon_l(\hat{p}, -) \epsilon_m(\hat{p}, -). 
\label{2-}
\end{equation}
In terms of spinors we will have 

\begin{equation}
e^{\ell n}_{mj}(\hat{p}, +2)= \frac{(-\imath\sigma_2)_{m r}\xi_p^\ell \xi_p^{ r}(-\imath\sigma_2)_{j s}\xi_p^n\xi_p^{ s}}{4|\vec{p}|^2}\,\, ,
 \end{equation}

\begin{equation}
e^{\ell n}_{mj}(\hat{p}, -2) = \frac{(-\imath\sigma_2)^{\ell r}\xi_{p r}\xi_{p m}(-\imath\sigma_2)^{n s}\xi_{p s}\xi_{p j}}{4|\vec{p}|^2}\,\, .
\end{equation}

The sum on the polarizations for gravitons $\Pi_{ij;kl}(\hat{q})$ amounts to (\ref{sumpol}). 

\section{Amplitudes and double copy }
\subsection{Tensor helicity amplitudes of correlator $\zeta\gamma\gamma$ }

Back to our correlator

\begin{eqnarray}
&&e_{ij}(\vec{p}_2,\lambda)G(\vec{p}_1, \vec{p}_2,\vec{p}_3)_{ij;lm}e_{lm}(\vec{p}_3,\lambda')=\nonumber\\ 
&&\delta^3 (\vec{p}_1+\vec{p}_2+\vec{p}_3)G_1(p_1, p_2, p_3)
(8\vec{p}_2\cdot\vec{p}_3)e_{ij}(\hat{p}_2,\lambda) e_{ij}(\hat{p}_3,\lambda').
\end{eqnarray}

The following products are needed

\begin{equation}
e_{ij}(\hat{p}_2,+2)e_{ij}(\hat{p}_3,+2)= \frac{(\xi_{p_2}^j(-\imath\sigma_2)_{j\ell}\xi_{p_3}^\ell)^4}{16 |\vec{p_2}|^2|\vec{p_3}|^2}\,\, ,
\label{ampl1}
\end{equation} 

\begin{equation}
e_{ij}(\hat{p}_2,-2)e_{ij}(\hat{p}_3,-2)= \frac{(\xi_{p_2\,j}(-\imath\sigma_2)^{j\ell}\xi_{p_3 \,\ell})^4}{16 |\vec{p_2}|^2|\vec{p_3}|^2}\,\, ,
\label{ampl2}
\end{equation} 

\begin{eqnarray}
&&e_{ij}(\hat{p}_2,-2)e_{ij}(\hat{p}_3,+2)= \frac{(\xi_{p_2\,j}{\xi_{p_3}}^j)^4}{16 |\vec{p_2}|^2 |\vec{p_3}|^2}\,\, ,
\label{ampl3}
\end{eqnarray} 

\begin{eqnarray}
&& e_{ij}(\hat{p}_2,+2)e_{ij}(\hat{p}_3,-2)= \frac{(\xi_{p_3\,j}{\xi_{p_2}}^j)^4}{16 |\vec{p_2}|^2 |\vec{p_3}|^2}\,\, .
\label{ampl4}
\end{eqnarray} 

\subsection{Double copy}\label{DC}

We propose that the amplitudes are the square of the flat space of type {\it Higgs to two gluons} amplitudes, based on a Lagrangian interaction in four dimensional Minkowski space $L_I \sim h F^2 $, where $F$ is the Yang- Mills field strength. This was noted in the context of correlators in conformal field theories in three dimensions in \cite{Farrow}. 
Let's just compare in fact the tensor factors. The corresponding color ordered amplitudes in four dimensional flat space are 
\begin{equation}
M(p_1; p_2,\lambda;p_3,\lambda ')\equiv p_2\cdot p_3 \epsilon(p_2,\lambda)\cdot \epsilon(p_3,\lambda') - p_2\cdot \epsilon(p_3,\lambda')p_3\cdot \epsilon(p_2,\lambda),
\label{higgs}
\end{equation}   
where $p$ stays for a four vector and the momentum conservation is implicit in (\ref{higgs}).
By relating the three dimensional polarization vectors to those of the four dimensional scattering amplitudes by
\begin{equation}
\epsilon^\mu (\vec{p}_i,\lambda_i)=\left(0, \vec{\epsilon}(\vec{p}_i,\lambda_i)\right)
\end{equation}
and the momenta so that
\begin{equation}
p_i^\mu= (p_i, \vec{p}_i)
\end{equation}
we obtain for the scattering amplitudes (\ref{higgs})
\begin{equation}
-(|\vec{p}_2||\vec{p}_3| - \vec{p}_2\cdot \vec{p}_3) \vec{ \epsilon}(\vec{p}_2,\lambda)\cdot \vec{\epsilon}(\vec{p}_3,\lambda')-\vec{p}_2\cdot \vec{\epsilon}(\vec{p}_3,\lambda')\vec{p}_3\cdot \vec{\epsilon}(\vec{p}_2,\lambda).
\end{equation}

Specializing at the different helicities we get the root square of (\ref{ampl4}), up to the factor $2\vec{p}_2\cdot \vec{p}_3$, 

\begin{eqnarray}
&& -(|\vec{p}_2||\vec{p}_3| - \vec{p}_2\cdot \vec{p}_3) \vec{ \epsilon}(\vec{p}_2,+)\cdot \vec{\epsilon}(\vec{p}_3,-)-\vec{p}_2\cdot \vec{\epsilon}(\vec{p}_3,-)\vec{p}_3\cdot \vec{\epsilon}(\vec{p}_2,+)=\\ \nonumber &&\frac{2\vec{p}_2\cdot \vec{p}_3}{4|\vec{p}_2||\vec{p}_3|} (\xi_{p_3}, \xi_{p_2})^2;
\label{Ampl1}
\end{eqnarray}
analogously for the root square of  (\ref{ampl1})
\begin{eqnarray}
&&-(|\vec{p}_2||\vec{p}_3| - \vec{p}_2\cdot \vec{p}_3) \vec{ \epsilon}(\vec{p}_2,+)\cdot \vec{\epsilon}(\vec{p}_3,+)-\vec{p}_2\cdot \vec{\epsilon}(\vec{p}_3,+)\vec{p}_3\cdot \vec{\epsilon}(\vec{p}_2,+)=\\ \nonumber
&& \frac{2\vec{p}_2\cdot \vec{p}_3}{4|\vec{p}_2||\vec{p}_3|} \left(\xi_{p_2}^i (-\imath\sigma_2)_{i\ell}\xi_{p_3}^\ell  \right)^2;
\label{Ampl2}
\end{eqnarray}
and finally for the root square of (\ref{ampl2})
\begin{eqnarray}
&&-(|\vec{p}_2||\vec{p}_3| - \vec{p}_2\cdot \vec{p}_3) \vec{ \epsilon}(\vec{p}_2,-)\cdot \vec{\epsilon}(\vec{p}_3,-)-\vec{p}_2\cdot \vec{\epsilon}(\vec{p}_3,-)\vec{p}_3\cdot \vec{\epsilon}(\vec{p}_2,-)=\nonumber\\
&& \frac{2\vec{p}_2\cdot \vec{p}_3}{4|\vec{p}_2||\vec{p}_3|} \left(\xi_{p_2\, i} (-\imath\sigma_2)^{i\ell}\xi_{p_3\,\ell}  \right)^2.
\label{Ampl3}
\end{eqnarray}
We conclude that
\begin{eqnarray}
&&e_{ij}(\vec{p}_2,\lambda)G(\vec{p}_1, \vec{p}_2,\vec{p}_3)_{ij;lm}e_{lm}(\vec{p}_3,\lambda')= 4\delta^3 (\vec{p}_1+\vec{p}_2+\vec{p}_3)\nonumber\\
&&\frac{G_1 (p_1,p_2,p_3)}{2\vec{p}_2\cdot \vec{p}_3} M^2\left(\vec{p}_1; \vec{p}_2,\frac{\lambda}{2};\vec{p}_3,\frac{\lambda '}{2}\right).
\label{DC1}
\end{eqnarray}
The three point correlator $\zeta\gamma\gamma$ is in its tensor part double copy of a gauge interaction of Higgs-like scalar coupled to gluons (\ref{higgs}). This is analogous to what was noticed in \cite{Farrow} for which in $\Phi R^2$ gravity the amplitude $graviton-graviton-scalar$ is the square in the flat space limit  of a correlator of two gauge currents and a marginal scalar. The compatibility with the result of \cite{Farrow} is due to their flat space limit and here has an analogous in the factorization of a form factor which dresses the Feynman diagrams  entering in the Minkowski space amplitude. The relation with gravity amplitudes found in (\ref{DC1}) is analogous to the KLT relations \cite{KLT}, but differing by a kinematic factor in the denominator. That factor allows the matching of the two spatial derivatives of the Ricci scalar intrinsic curvature $R^{(3)}=g_{ij}R^{(3)}_{ij}$ and the four derivatives appearing in the square of $h F^2$ on the gauge side.

\section{A one-loop calculation}

For the one-loop contribution to the correlation function of two $\zeta$ fields, which is measured in the spectrum of anisotropies of the cosmic microwave background, we are going to explore how to use the proved double copy relation (\ref{DC1}).  
Consider
\begin{equation}
C(\vec{q})=\int d^ 3x e^{\imath\vec{q}\cdot (\vec{x}-\vec{x}\,')}<\zeta(\vec{x},t)\zeta(\vec{x}\,',t)>_{2\gamma}
\end{equation}

where the subscript $2\gamma$ means that we are calculating $G(q)$ at the second order in the interaction by the exchange of two gravitational waves. The time dependent Hamiltonian to be used in the In-In formalism is (\ref{zgg}). The actual contribution will be given by the part of the Hamiltonian which is not a total derivative, meaning

\begin{equation}
A(t)= -\frac{\epsilon a}{8}\int d^3 x \zeta(\vec{x},t)\partial_m\gamma_{ri}\partial_m\gamma_{ri}(\vec{x},t).
\label{A1}
\end{equation}

By computing 
\begin{equation}
-\int d^ 3x e^{\imath\vec{q}\cdot (\vec{x}-\vec{x}\,')}\int_{-\infty}^t dt_2\int_{-\infty}^{t_2} dt_1<[A(t_1),[A(t_2),\zeta(\vec{x},t)\zeta(\vec{x}\,',t)]] >\nonumber\\
\end{equation}

we find

\begin{eqnarray}
C(\vec{q})&=& -\frac{1}{8}\Re \int_{-\infty}^t dt_2\int_{-\infty}^{t_2} dt_1 (2\pi)^9\\ \nonumber
&& \frac{\epsilon(t_1)a(t_1)}{4}\frac{\epsilon(t_2)a(t_2)}{4}\left(\zeta_q(t_2)\zeta_q(t)^*- \zeta_q(t)\zeta_q(t_2)^*\right)\zeta_q(t_1)\zeta_q(t)^*\nonumber\\
&& \int d^3 q_2\int d^3 q_3 \delta^3(\vec{q}_2+\vec{q}_3+\vec{q})(2\vec{q}_2\cdot\vec{q}_3)^2 \Pi_{ri;r_1i_1}(\hat{q}_2)\Pi_{ri;r_1i_1}(\hat{q}_3)\nonumber\\
&&\gamma_{q_2}(t_1)\gamma_{q_2}(t_2)^* \gamma_{q_3}(t_1)\gamma_{q_3}(t_2)^* 
\end{eqnarray}

where $\Re$ stays for the real part.

\subsection{Observations}
\begin{enumerate}
\item The asymmetry between the two vertices implies a different treatment of the external legs. Actually in \cite{Wein}, \cite{Giddings} it is distinguished between the left and right vertices with which the external legs have different propagators, we prefer here an operator approach more than diagrammatic.
\item The loop integral 
\begin{eqnarray}
&&I= \int d^3 q_2\int d^3 q_3 \delta^3(\vec{q}+\vec{q}_2+\vec{q}_3)(2\vec{q}_2\cdot\vec{q}_3)^2 \Pi_{ri;r_1i_1}(\hat{q}_2)\Pi_{ri;r_1i_1}(\hat{q}_3)\nonumber\\
&&\gamma_{q_2}(t_1)\gamma_{q_2}(t_2)^* \gamma_{q_3}(t_1)\gamma_{q_3}(t_2)^* 
\end{eqnarray}
has the analogous on the denominators of the Feynman propagators in the products $\gamma_{q_2}(t_1)\gamma_{q_2}(t_2)^*$. The tensor part instead plays the role of the numerators of the Feynman diagrams for scattering amplitudes.

\item The tensor factor 
\begin{equation}
(2\vec{q}_2\cdot\vec{q}_3)^2 \Pi_{ri;r_1i_1}(\hat{q}_2)\Pi_{ri;r_1i_1}(\hat{q}_3)
\end{equation}
amounts to
\begin{equation}
\sum_{\lambda,\lambda'=\pm1}\frac{M^2(\vec{q};\vec{q}_2,\lambda;\vec{q}_3,\lambda ')M^{2}(-\vec{q}; -\vec{q}_2,-\lambda;-\vec{q}_3,-\lambda ')}{(2\vec{q}_2\cdot \vec{q}_3)^2}
\end{equation}
where $M(\vec{q},\vec{q}_2,\lambda;\vec{q}_3,\lambda ')$ is defined as (\ref{higgs}). We therefore explicitly verify that combining in a bigger diagram trilinear vertices, non containing time derivatives, the resulting tensor part is the product of the tensor factors \cite{kltINFLA}. 
\end{enumerate}

\section{Four points tree level correlator $<\zeta\zeta\gamma\gamma>$}
\subsection{The $t$ and $u$ channels}
We are going to explore double copy in a sort of primordial trispectrum with two scalars and two tensor fluctuactions
\begin{eqnarray}
&& G(\vec{p}_1,\vec{p}_2,\vec{p}_3,\vec{p}_4)_{ij;\ell m}\equiv <\int d^3x_1 d^3x_2 d^3x_3 d^3x_4 \exp{\imath\vec{p}_1\cdot \vec{x}_1} \exp{\imath\vec{p}_2\cdot \vec{x}_2} \nonumber\\
&&\exp{\imath\vec{p}_3\cdot \vec{x}_3} \exp{\imath\vec{p}_4\cdot \vec{x}_4}\,\zeta(\vec{x}_1,t)\zeta(\vec{x}_2,t)\gamma_{ij}(\vec{x}_3,t)\gamma_{\ell m}(\vec{x}_4,t)>.
\label{G4}
\end{eqnarray}
We are going just to consider cubic vertices because higher order interactions are irrelevant in the sense of the renormalization flow.  We will discover in the different channels that the tensor parts of the given correlator with three tensor fluctuactions, two scalars and one tensor fluctuactions and finally two tensors and one scalar fluctuactions will be building blocks of that trispectrum.  
The second order in (\ref{A1}), which involves cubic vertices and provides the $t$ and $u$ channels with one $\gamma$ exchange, amounts to
{\allowdisplaybreaks
\begin{eqnarray}
&& e_{ij}(\vec{p}_3,\lambda) G(\vec{p}_1,\vec{p}_2,\vec{p}_3,\vec{p}_4)_{ij;\ell m} e_{\ell m}(\vec{p}_4,\lambda')=(2\pi)^{18}\delta^3(\vec{p}_1+\vec{p}_2+\vec{p}_3+ \vec{p}_4)\times\nonumber\\
&&\left[\Re \int_{-\infty}^t dt_2\frac{\epsilon(t_2)a(t_2)}{4} \int_{-\infty}^{t_2} dt_1\frac{\epsilon(t_1)a(t_1)}{4} \left(\zeta_{p_1}(t_2)\zeta_{p_1}(t)^*- \zeta_{p_1}(t)\zeta_{p_1}(t_2)^*\right)\right. \nonumber\\
&& \left.\zeta_{p_2}(t_1)\zeta_{p_2}(t)^*\gamma_{p_3}(t_1)\gamma_{p_3}(t)^*\gamma_{p_4}(t_2)\gamma_{p_4}(t)^* \gamma_{|\vec{p}_4 +\vec{p}_1|}(t_1) \gamma^*_{|\vec{p}_4 +\vec{p}_1|}(t_2)\right.\nonumber\\
&&\left. \sum_{\lambda_1=\pm 1}\frac{M^2(\vec{p}_1,\vec{p}_4,\lambda';- \vec{p}_1-\vec{p}_4,\lambda_1)M^2(\vec{p}_2,\vec{p}_3,\lambda; \vec{p}_1+\vec{p}_4,-\lambda_1)}{(2\vec{p}_4\cdot({\vec{p}_1+\vec{p}_4}))^2}\right.\nonumber\\
&&\left.+\Re \int_{-\infty}^t dt_2\frac{\epsilon(t_2)a(t_2)}{4} \int_{-\infty}^{t_2} dt_1\frac{\epsilon(t_1)a(t_1)}{4} \left(\zeta_{p_1}(t_2)\zeta_{p_1}(t)^*- \zeta_{p_1}(t)\zeta_{p_1}(t_2)^*\right)\right.\nonumber\\
&&\left. \zeta_{p_2}(t_1)\zeta_{p_2}(t)^*\gamma_{p_3}(t_2)\gamma_{p_3}(t)^*\gamma_{p_4}(t_1)\gamma_{p_4}(t)^* \right.\gamma_{|\vec{p}_1 +\vec{p}_3|}(t_1) \gamma^ *_{|\vec{p}_1 +\vec{p}_3|}(t_2)\nonumber\\
&& \left. \sum_{\lambda_1=\pm 1}\frac{M^2(\vec{p}_1,\vec{p}_3,\lambda ; -\vec{p}_1-\vec{p}_3,\lambda_1)M^2(\vec{p}_2,\vec{p}_4,\lambda';\vec{p}_1+\vec{p}_3,-\lambda_1)}{(2\vec{p}_3\cdot({\vec{p}_1+\vec{p}_3}))^2} \right.\nonumber\\
&&\left.+\Re \int_{-\infty}^t dt_2\frac{\epsilon(t_2)a(t_2)}{4} \int_{-\infty}^{t_2} dt_1\frac{\epsilon(t_1)a(t_1)}{4}\left(\zeta_{p_2}(t_2)\zeta_{p_2}(t)^*- \zeta_{p_2}(t)\zeta_{p_2}(t_2)^*\right) \right.\nonumber\\
&& \left.\zeta_{p_1}(t_1)\zeta_{p_1}(t)^*\gamma_{p_3}(t_1)\gamma_{p_3}(t)^*\gamma_{p_4}(t_2)\gamma_{p_4}(t)^* \gamma_{|\vec{p}_2 +\vec{p}_4|}(t_1) \gamma^ *_{|\vec{p}_2 +\vec{p}_4|}(t_2)
\right.\nonumber\\
&&\left.\sum_{\lambda_1=\pm 1} \frac{M^2(\vec{p}_2,\vec{p}_4,\lambda' ; -\vec{p}_2-\vec{p}_4,\lambda_1)M^2(\vec{p}_1,\vec{p}_3,\lambda; \vec{p}_2+\vec{p}_4,-\lambda_1)}{(2\vec{p_4}\cdot({\vec{p}_2+\vec{p}_4}))^2}\right.\nonumber\\
&&\left.+\Re \int_{-\infty}^t dt_2\frac{\epsilon(t_2)a(t_2)}{4} \int_{-\infty}^{t_2} dt_1\frac{\epsilon(t_1)a(t_1)}{4}\left(\zeta_{p_2}(t_2)\zeta_{p_2}(t)^*- \zeta_{p_2}(t)\zeta_{p_2}(t_2)^*\right) \right.\nonumber\\
&& \left.\zeta_{p_1}(t_1)\zeta_{p_1}(t)^*\gamma_{p_3}(t_2)\gamma_{p_3}(t)^*\gamma_{p_4}(t_1)\gamma_{p_4}(t)^*\gamma_{|\vec{p}_2 +\vec{p}_3|}(t_1) \gamma^ *_{|\vec{p}_2 +\vec{p}_3|}(t_2)
 \right.\nonumber\\
&&\left.\sum_{\lambda_1=\pm 1}\frac{M^2(\vec{p}_2,\vec{p}_3,\lambda ; -\vec{p}_2-\vec{p}_3,\lambda_1)M^2(\vec{p}_1,\vec{p}_4,\lambda'; \vec{p}_2+\vec{p}_3,-\lambda_1)}{{(2\vec{p}_3 \cdot (\vec{p}_2+\vec{p}_3))^2}}\right.\nonumber\\
&&\left.+\Re \int_{-\infty}^t dt_2\frac{\epsilon(t_2)a(t_2)}{4} \int_{-\infty}^{t_2} dt_1\frac{\epsilon(t_1)a(t_1)}{4}  \left(\gamma_{p_3}(t_2)\gamma_{p_3}(t)^*- \gamma_{p_3}(t)\gamma_{p_3}(t_2)^*\right) \right.\nonumber\\
&&\left.\zeta_{p_1}(t_2)\zeta_{p_1}(t)^*\zeta_{p_2}(t_1)\zeta_{p_2}(t)^*\gamma_{p_4}(t_1)\gamma_{p_4}(t)^* \gamma_{|\vec{p}_1 +\vec{p}_3|}(t_1) \gamma^ *_{|\vec{p}_1 +\vec{p}_3|}(t_2)
\right.\nonumber\\
&& \left.\sum_{\lambda_1=\pm 1}\frac{M^2(\vec{p}_1,\vec{p}_3,\lambda ; -\vec{p}_1-\vec{p}_3,\lambda_1)M^2(\vec{p}_2,\vec{p}_4,\lambda'; \vec{p}_1+\vec{p}_3,-\lambda_1)}{{(2\vec{p}_3 \cdot (\vec{p}_1+\vec{p}_3))^2}}\right.\nonumber\\
&&\left.+\Re \int_{-\infty}^t dt_2\frac{\epsilon(t_2)a(t_2)}{4} \int_{-\infty}^{t_2} dt_1\frac{\epsilon(t_1)a(t_1)}{4}  \left(\gamma_{p_3}(t_2)\gamma_{p_3}(t)^*- \gamma_{p_3}(t)\gamma_{p_3}(t_2)^*\right)\right.\nonumber\\
&&\left. \zeta_{p_1}(t_1)\zeta_{p_1}(t)^*\zeta_{p_2}(t_2)\zeta_{p_2}(t)^*\gamma_{p_4}(t_1)\gamma_{p_4}(t)^* \gamma_{|\vec{p}_2 +\vec{p}_3|}(t_1) \gamma^ *_{|\vec{p}_2 +\vec{p}_3|}(t_2)+
\right.\nonumber\\
&&\left. \sum_{\lambda_1=\pm 1}\frac{M^2(\vec{p}_2,\vec{p}_3,\lambda ; -\vec{p}_2-\vec{p}_3,\lambda_1)M^2(\vec{p}_1,\vec{p}_4,\lambda'; \vec{p}_2+\vec{p}_3,-\lambda_1)}{(2\vec{p}_3 \cdot (\vec{p}_2+\vec{p}_3))^2}\right.\nonumber\\
&&\left.+\Re \int_{-\infty}^t dt_2\frac{\epsilon(t_2)a(t_2)}{4} \int_{-\infty}^{t_2} dt_1\frac{\epsilon(t_1)a(t_1)}{4} \left(\gamma_{p_4}(t_2)\gamma_{p_4}(t)^*- \gamma_{p_4}(t)\gamma_{p_4}(t_2)^*\right) \right.\nonumber\\
&&\left. \zeta_{p_1}(t_1)\zeta_{p_1}(t)^*\zeta_{p_2}(t_2)\zeta_{p_2}(t)^*\gamma_{p_3}(t_1)\gamma_{p_3}(t)^* \gamma_{|\vec{p}_2 +\vec{p}_4|}(t_1) \gamma^ *_{|\vec{p}_2 +\vec{p}_4|}(t_2)
\right.\nonumber\\
&& \left.\sum_{\lambda_1=\pm 1}\frac{M^2(\vec{p}_2,\vec{p}_4,\lambda' ; -\vec{p}_2-\vec{p}_4,\lambda_1)M^2(\vec{p}_1,\vec{p}_3,\lambda; \vec{p}_2+\vec{p}_4,-\lambda_1)}
{(2\vec{p}_4 \cdot (\vec{p}_2+\vec{p}_4))^2}\right.\nonumber\\&&
\left.+\Re \int_{-\infty}^t dt_2\frac{\epsilon(t_2)a(t_2)}{4} \int_{-\infty}^{t_2} dt_1\frac{\epsilon(t_1)a(t_1)}{4}  \left(\gamma_{p_4}(t_2)\gamma_{p_4}(t)^*- \gamma_{p_4}(t)\gamma_{p_4}(t_2)^*\right)\right.\nonumber\\
&&\left. \zeta_{p_1}(t_2)\zeta_{p_1}(t)^*\zeta_{p_2}(t_1)\zeta_{p_2}(t)^*\gamma_{p_3}(t_1)\gamma_{p_3}(t)^* \gamma_{|\vec{p}_1 +\vec{p}_4|}(t_1) \gamma^ *_{|\vec{p}_1 +\vec{p}_4|}(t_2)
\right.\nonumber\\
&&\left. \sum_{\lambda_1= \pm 1}\frac{M^2(\vec{p}_1,\vec{p}_4,\lambda' ; -\vec{p}_1-\vec{p}_4,\lambda_1)M^2(\vec{p}_2,\vec{p}_3,\lambda; \vec{p}_1+\vec{p}_4,-\lambda_1)}{(2\vec{p}_4\cdot (\vec{p}_1+\vec{p}_4))^2}\right]\label{quartic1}
\end{eqnarray}
}
All possible permutations of the four external legs are included and again $M(\vec{p},\vec{q},\lambda' ; -\vec{p}-\vec{q},\lambda_1)$ is defined in (\ref{higgs}).

\subsection{The $s$ channel}

The tree level $s$ channel of $e_{ij}(\hat{p}_3,\lambda_3) G(\vec{p}_1,\vec{p}_2,\vec{p}_3,\vec{p}_4)_{ij;\ell m} e_{\ell m}(\hat{p}_4,\lambda_4)$ receives the lowest order contribution from the following terms in the Hamiltonian
\begin{equation}
H_I(t)=\tilde{A}_1(t) +\tilde{A}_2(t)
\end{equation}
where
\begin{equation}
\tilde{A}_1(t)= \int d^3 x a(t)\left[\gamma_{ij}(\vec{x},t)\partial_i\zeta(\vec{x},t)\partial_j\zeta(\vec{x},t)+2\gamma_{ij}(\vec{x},t)\zeta(\vec{x},t)\partial_i\partial_j \zeta(\vec{x},t)\right]
\end{equation}
\begin{equation}
\tilde{A}_2(t) = -\frac{1}{8}\int d^3 x a(t)\left[\gamma_{kl}(\vec{x},t)\partial_k\gamma_{ij}(\vec{x},t)-2\gamma_{ik}(\vec{x},t)\partial_k\gamma_{jl}(\vec{x},t)\right]\partial_l\gamma_{ij}(\vec{x},t).
\end{equation}
Only the contributions from one $\gamma$ exchange will be taken into account, therefore the claimed $s-$channel amounts to
{\allowdisplaybreaks
\begin{eqnarray}
&&S=-{2}\int_{-\infty}^ t dt_2\int_{-\infty}^{t_2} dt_1 a(t_1) a(t_2)(2\pi)^{18}\delta^{3}(\vec{p}_1+\vec{p}_2 + \vec{p}_3+ \vec{p}_4 )\times\nonumber\\
&&\left\{\zeta_{p_2}^*(t_2)\zeta_{p_2}(t)\gamma_{p_3}(t_1)\gamma^*_{p_3}(t)\gamma_{p_4}(t_1)\gamma^*_{p_4}(t)
(\zeta_{p_1}(t_2)\zeta^*_{p_1}(t)- \zeta^*_{p_1}(t_2)\zeta_{p_1}(t))\times\nonumber\right.\\
&&\left.(\gamma_{|\vec{p}_3+\vec{p}_4|}(t_1)\gamma_{|\vec{p}_3+\vec{p}_4|}^*(t_2)- \gamma_{|\vec{p}_3+\vec{p}_4|}(t_2)\gamma_{|\vec{p}_3+\vec{p}_4|}^*(t_1))\right.\nonumber\\
&& +\zeta_{p_1}^*(t_2)\zeta_{p_1}(t)\gamma_{p_3}(t_1)\gamma^*_{p_3}(t)\gamma_{p_4}(t_1)\gamma^*_{p_4}(t)(\zeta_{p_2}(t_2)\zeta^*_{p_2}(t)- \zeta^*_{p_2}(t_2)\zeta_{p_2}(t))\nonumber\\
&&(\gamma_{|\vec{p}_3+\vec{p}_4|}(t_1)\gamma_{|\vec{p}_3+\vec{p}_4|}^*(t_2)- \gamma_{|\vec{p}_3+\vec{p}_4|}(t_2)\gamma_{|\vec{p}_3+\vec{p}_4|}^*(t_1))+ (\zeta_{p_1}(t_2)\zeta^*_{p_1}(t)- \zeta^*_{p_1}(t_2)\zeta_{p_1}(t))\times \nonumber\\
&&\left[\gamma_{p_4}(t_1)\gamma_{p_3}(t_1)\gamma_{|\vec{p}_3+\vec{p}_4|}(t_2)\zeta_{p_2}(t)\gamma^*_{p_4}(t)\gamma^*_{p_3}(t)\gamma^*_{|\vec{p}_3+\vec{p}_4|}(t_1)\zeta^*_{p_2}(t_2)\right.\nonumber\\
&&\left.- \gamma_{p_4}(t)\gamma_{p_3}(t)\gamma_{|\vec{p}_3+\vec{p}_4|}(t_2)\zeta_{p_2}(t)\gamma^*_{p_4}(t_1)\gamma^*_{p_3}(t_1)\gamma^*_{|\vec{p}_3+\vec{p}_4|}(t_1)\zeta^*_{p_2}(t_2)\right]\nonumber\\
&&+ (\zeta_{p_2}(t_2)\zeta^*_{p_2}(t)- \zeta^*_{p_2}(t_2)\zeta_{p_2}(t))\nonumber\\
&&\left[\gamma_{p_4}(t_1)\gamma_{p_3}(t_1)\gamma_{|\vec{p}_3+\vec{p}_4|}(t_2)\zeta_{p_1}(t)\gamma^*_{p_4}(t)\gamma^*_{p_3}(t)\gamma^*_{|\vec{p}_3+\vec{p}_4|}(t_1)\zeta^*_{p_1}(t_2)\right.\nonumber\\
&&\left.- \gamma_{p_4}(t)\gamma_{p_3}(t)\gamma_{|\vec{p}_3+\vec{p}_4|}(t_2)\zeta_{p_1}(t)
\gamma^*_{p_4}(t_1)\gamma^*_{p_3}(t_1)\gamma^*_{|\vec{p}_3+\vec{p}_4|}(t_1)\zeta^*_{p_1}(t_2)\right]+\nonumber\\
&&\left(\gamma_{p_3}(t_2)\gamma^*_{p_3}(t)-\gamma^*_{p_3}(t_2)\gamma_{p_3}(t)\right) (\gamma_{|\vec{p}_3+\vec{p}_4|}(t_1)\gamma_{|\vec{p}_3+\vec{p}_4|}^*(t_2)- \gamma_{|\vec{p}_3+\vec{p}_4|}(t_2)\gamma_{|\vec{p}_3+\vec{p}_4|}^*(t_1))\nonumber\\
&&\zeta_{p_1}(t_1)\zeta_{p_2}(t_1)\gamma_{p_4}(t_2)\zeta_{p_1}^*(t_)\zeta_{p_2}^*(t)\gamma_{p_4}^*(t) + \left(\gamma_{p_4}(t_2)\gamma^*_{p_4}(t)-\gamma^*_{p_4}(t_2)\gamma_{p_4}(t)\right)\nonumber\\
&& (\gamma_{|\vec{p}_3+\vec{p}_4|}(t_1)\gamma_{|\vec{p}_3+\vec{p}_4|}^*(t_2)- \gamma_{|\vec{p}_3+\vec{p}_4|}(t_2)\gamma_{|\vec{p}_3+\vec{p}_4|}^*(t_1))\times\nonumber\\&&\left.\zeta_{p_1}(t_1)\zeta_{p_1}^*(t)\zeta_{p_2}(t_1)\zeta_{p_2}^*(t)\gamma_{p_3}(t_2)\gamma_{p_3}^*(t)\right\} p_{1i_1}p_{1j_1}\sum_\lambda e_{i_1j_1}(\vec{p}_1+\vec{p}_2,\lambda) e_{i_2 j_2} (-\vec{p}_3-\vec{p}_4, -\lambda)\nonumber\\&&\left [ p_{4 i}e_{ij}(\vec{p}_3,\lambda_3)p_{4 j} e_{i_2 j_2}(\vec{p}_4,\lambda_4)+p_{3 i}e_{ij}(\vec{p}_4,\lambda_4)p_{4 j}e_{i_2 j_2}(\vec{p}_3,\lambda_3)-p_{3 i_2}p_{4 j_2}e_{ij}(\vec{p}_3,\lambda_3) e_{ij}(\vec{p}_4,\lambda_4)\right.\nonumber\\
&& \left. -2p_{4 i_2}e_{j_2 l_1}(\vec{p}_4,\lambda_4)e_{l_1 k}(\vec{p}_3,\lambda_3) p_{4 k} - 2p_{3 i_2}e_{j_2 l_1 }(\vec{p}_3,\lambda_3)e_{l_1 k}(\vec{p}_4,\lambda_4) p_{3 k}\right.\nonumber\\ 
&&\left. - 2p_{3 l_1}e_{l_1 j_2}(\vec{p}_3,\lambda_3)e_{i_2 k}(\vec{p}_4,\lambda_4) p_{3 k} \right]\nonumber\\ 
\label{S}
\end{eqnarray}
}
The tensor part of $S$ on which we are going to focus in order to recognize a double copy structure is
\begin{eqnarray}
&& p_{1i_1}p_{1j_1}\sum_\lambda e_{i_1j_1}(\vec{p}_1+\vec{p}_2,\lambda)e_{i_2 j_2} (-\vec{p}_3-\vec{p}_4, -\lambda)\left [ p_{4 i}e_{ij}(\vec{p}_3,\lambda_3)p_{4 j} e_{i_2 j_2}(\vec{p}_4,\lambda_4)\right.\nonumber\\
&&\left.+p_{3 i}e_{ij}(\vec{p}_4,\lambda_4)p_{4 j}e_{i_2 j_2}(\vec{p}_3,\lambda_3)-p_{3 i_2}p_{4 j_2}e_{ij}(\vec{p}_3,\lambda_3) e_{ij}(\vec{p}_4,\lambda_4)-2p_{4 i_2}e_{j_2 l_1}(\vec{p}_4,\lambda_4)\times\right.\nonumber\\
&&\left. e_{l_1 k}(\vec{p}_3,\lambda_3) p_{4 k} - 2p_{3 i_2}e_{j_2 l_1 }(\vec{p}_3,\lambda_3)e_{l_1 k}(\vec{p}_4,\lambda_4) p_{3 k} - 2p_{3 l_1}e_{l_1 j_2}(\vec{p}_3,\lambda_3)e_{i_2 k}(\vec{p}_4,\lambda_4) p_{3 k} \right].\nonumber\\ 
\label{T}
\end{eqnarray}
Concerning the remaining scalar part of (\ref{S}) we consider to write it in that form for compactness because no significative simplifications arise by expanding it.  

By the prescription of the polarization tensors for the gravitons as defined in (\ref{2+}) and (\ref{2-}) we may rewrite one of the two factors in the helicity sum of (\ref{T}) as 

\begin{eqnarray}
&&\left(\vec{p}_4\cdot \vec{\epsilon}(\vec{p}_3)\right)^2\left(\vec{\epsilon}(\vec{p}_4)\cdot \vec{\epsilon}(-\vec{p}_3-\vec{p}_4)\right)^2 + \left(\vec{p}_3\cdot \vec{\epsilon}(\vec{p}_4)\right)^2\left(\vec{\epsilon}(\vec{p}_3)\cdot \vec{\epsilon}(-\vec{p}_3-\vec{p}_4)\right)^2\nonumber\\ 
&& + \left(\vec{p}_3\cdot \vec{\epsilon}(-\vec{p}_3-\vec{p}_4)\right)^2\left(\vec{\epsilon}(\vec{p}_3)\cdot \vec{\epsilon}(\vec{p}_4)\right)^2 \nonumber\\
&& -2\vec{p}_4\cdot\vec{\epsilon}(\vec{p}_4) \vec{\epsilon}(\vec{p}_4)\cdot \vec{\epsilon}(\vec{p}_3)  \vec{\epsilon}(\vec{p}_3)\cdot \vec{\epsilon}(-\vec{p}_3-\vec{p}_4) \vec{\epsilon}(-\vec{p}_3-\vec{p}_4)\cdot \vec{\epsilon}(\vec{p}_4) \vec{\epsilon}(\vec{p}_4)\cdot \vec{p}_3\nonumber\\
&& +2 \vec{p}_3\cdot\vec{\epsilon}(\vec{p}_4)\vec{\epsilon}(\vec{p}_4)\cdot \vec{\epsilon}(\vec{p}_3)\vec{\epsilon}(\vec{p}_3)\cdot \vec{\epsilon}(-\vec{p}_3- \vec{p}_4)  \vec{\epsilon}(-\vec{p}_3- \vec{p}_4)\cdot \vec{p}_4\nonumber\\
&& +2 \vec{p}_3\cdot\vec{\epsilon}(-\vec{p}_4-\vec{p}_3)\vec{\epsilon}(-\vec{p}_4-\vec{p}_3)\cdot\vec{\epsilon}(\vec{p}_4)\vec{\epsilon}(\vec{p}_3)\cdot \vec{\epsilon}(\vec{p}_4)  \vec{\epsilon}(\vec{p}_3)\cdot \vec{p}_4,
\label{[]}
\end{eqnarray}
with the label of the helicity not explicitly indicated. (\ref{[]}) amounts to the squared amplitude
\begin{eqnarray}
&&\left[ \vec{p}_4\cdot \vec{\epsilon}(\vec{p}_3) \vec{\epsilon}(\vec{p}_4)\cdot \vec{\epsilon}(-\vec{p}_3-\vec{p}_4) - \vec{p}_3\cdot \vec{\epsilon}(\vec{p}_4)\left(\vec{\epsilon}(\vec{p}_3)\cdot \vec{\epsilon}(-\vec{p}_3-\vec{p}_4)\right)\right.\nonumber\\
&&\left.+ \left(\vec{p}_3\cdot \vec{\epsilon}(-\vec{p}_3-\vec{p}_4)\right)\left(\vec{\epsilon}(\vec{p}_3)\cdot \vec{\epsilon}(\vec{p}_4)\right)\right]^2
\end{eqnarray}
that is the square of a color ordered amplitude in the renormalizable Yang-Mills theory.
The other factor 
\begin{eqnarray}
&&(p_{1i_2}p_{2j_2}+p_{1i_2}p_{1j_2}+ p_{2i_2}p_{2j_2} )\epsilon_{i_2j_2}(-\vec{p}_1-\vec{p}_2,\lambda)= p_{1i_2}p_{1j_2} \epsilon_{i_2j_2}(-\vec{p}_1-\vec{p}_2,\lambda)\nonumber\\
&& =(p_{1i}\epsilon_{i}(-\vec{p}_1-\vec{p}_2,\lambda))^2=\frac{1}{4}((p_1-p_2)_i \epsilon_{i}(-\vec{p}_1-\vec{p}_2,\lambda))^2
\end{eqnarray}
corresponds to the renormalizable coupling $scalar-scalar-gluon$ already remarked in a different context in \cite{Farrow}.


\section{Conclusions}

In the class of theories of inflation characterized by an inflaton scalar field we computed correlators that involve the curvature perturbation $\zeta$ as well as the gravitational wave amplitudes $\gamma_{ij}$. We have recognized that the building blocks of three point amplitudes in terms of which correlators of higher multiplicity could be constructed manifest in their tensor part double copy. The three point function $<\zeta\gamma\gamma>$, can be factorized into a time dependent scalar factor and a time independent so-called tensor factor which is recognized as squared of scattering amplitude of a gauge invariant interaction of type $scalar-vector-vector$ in the four dimensional flat space. By those blocks we may construct the tensor part of the one loop correlator $<\zeta\zeta>$. It is worth to remark that we have extended the equality $Gravity=(Gauge)^2$ \cite{BCJ}, since gauge does not mean just Yang-Mills  beacause we have included a gauge invariant interaction of a scalar with gluons of type $\sim h F^2$. The correlators are computed by the mean of the so called In-In formalism, however at one loop level we recognize an integrand structure in terms of loop momentum very similar to those of the Feynman integrals for scattering amplitudes. The analogous of the denominators is indeed given by the product of the modes satisfying the equations (\ref{MUK}), (\ref{GRAV}).The analogous of numerator of the Feynman integrals is given by the tensor part reproducing tree level scattering amplitudes. The In-In formalism by its construction seems to implement the on shell techniques together with the generalized unitarity methods to compute cosmological correlators in perturbation theory. We found that the numerators are products of three point amplitudes in flat space, so that one can recover the tensor part as product of the tensor factors. What we add here is that despite of the complications of the vertices of gravitational interactions we have the prescription that such constituents building blocks are square of gauge theory amplitudes from an interaction of type  $scalar-vector-vector$. 

Moreover the tree-level four point correlator $<\zeta\zeta\gamma\gamma>$ has been constructed. It is possible to recognize the channels of type $t$, $u$ and $s$ by graviton exchange. The terms entering in the tensor part are obtained as the product of three point amplitudes. For the channels $t$ and $u$, they express as the square of the amplitudes of an interaction of type $scalar-vector-vector$, instead for the $s$ channel we have two factors which are the square of $vector-vector-vector$ interaction in pure Yang-Mills theory and the other factor is instead related to the squared amplitude $scalar-scalar-vector$. This result is encouraging toward the identification of a double copy structure in a four point correlator. It happens that all the mentioned gauge interactions can be recast into a gauge theory marginal in six dimensions and recently constructed in \cite{Johansson} in order for the scattering amplitudes obeying color-kinematic duality so that to reproduce the amplitudes of conformal gravity. Actually the choice of disregarding the irrelevant interactions brought to consider only trilinear vertices however by involving higher multiplicity interactions also higher derivatives  Yang-Mills interactions can arise as in the formulation of  \cite{Johansson}.

Our present understanding of the way consistent time evolution is encoded in cosmological correlators is still in its infancy, matching not even the level of understanding
for tree-level scattering amplitudes. Here we make an attempt to see if double copy could emerge in correlators by the In-In formalism also motivated by \cite{kosower} . We have written cosmological correlators in a form that involves at integrand level gravitational building blocks that are the square of the gauge building blocks amplitudes. In the organization of  the calculations in the In-In formalism in a different form than that of Schwinger-Keldysh there are indications that the double copy could apply as well bootstrapp like techniques to reconstruct the integrands of cosmological correlators.

\section*{Acknowledgments}
The International Institute of Physics from the Federal University of Rio
Grande Do Norte (IIP-UFRN) is kindly acknowledged for the wonderful hospitality and generous financial support.


\end{document}